\documentclass[prl,reprint,twocolumn,preprintnumbers,amsmath,amssymb,superscriptaddress]{revtex4}
\usepackage[pdftex]{graphicx, color}
\usepackage{amsmath}
\usepackage{amssymb}
\usepackage{amsthm}
\usepackage{amsfonts}						%
\usepackage{graphicx}
\usepackage[english]{babel}             	
\usepackage{epstopdf}
\usepackage{graphicx}                   	
\usepackage{color}
\usepackage{SIunits}

\begin{document}

\title{Photonic-Crystal Waveguides with Disorder: Measurement of a Band-Edge Tail in the Density of States}

\author{S.R. Huisman}  \email{s.r.huisman@utwente.nl, www.utwente.nl/mesaplus/anp}
\affiliation{MESA+ Institute for
Nanotechnology, University of Twente, PO Box 217, 7500 AE Enschede, The Netherlands}
\author{G. Ctistis}
\affiliation{MESA+ Institute for
Nanotechnology, University of Twente, PO Box 217, 7500 AE Enschede, The Netherlands}
\author{S. Stobbe}
\affiliation{Niels Bohr Institute, University of Copenhagen, Blegdamsvej 17, DK-2100, Copenhagen, Denmark}
\author{A.P. Mosk}
\affiliation{MESA+ Institute for
Nanotechnology, University of Twente, PO Box 217, 7500 AE Enschede, The Netherlands}
\author{J.L. Herek}
\affiliation{MESA+ Institute for
Nanotechnology, University of Twente, PO Box 217, 7500 AE Enschede, The Netherlands}
\author{A. Lagendijk}
\affiliation{MESA+ Institute for
Nanotechnology, University of Twente, PO Box 217, 7500 AE Enschede, The Netherlands}
\affiliation{FOM Institute for Atomic and Molecular Physics, Science Park 104, 1098 XG Amsterdam, The Netherlands}
\author{P. Lodahl}
\affiliation{Niels Bohr Institute, University of Copenhagen, Blegdamsvej 17, DK-2100, Copenhagen, Denmark}
\author{W.L. Vos}
\affiliation{MESA+ Institute for
Nanotechnology, University of Twente, PO Box 217, 7500 AE Enschede, The Netherlands}
\author{P.W.H. Pinkse}
\affiliation{MESA+ Institute for
Nanotechnology, University of Twente, PO Box 217, 7500 AE Enschede, The Netherlands}

\date{\today}

\begin{abstract}

We measure localized and extended mode profiles at the band edge of slow-light photonic-crystal waveguides using phase-sensitive near-field microscopy.
High-resolution band structures are obtained and interpreted, allowing the retrieval of the optical density of states (DOS). This constitutes a first  observation of the DOS of a periodic system with weak disorder. The Van Hove singularity in the DOS expected at the band edge of an ideal 1D periodic structure is removed by the disorder. The Anderson-localized states form a ``tail'' in the density of states, as predicted by Lifshitz for solid-state systems.


\end{abstract}
\maketitle


Band gaps and accompanying band-edge effects are among the most-studied phenomena in solid-state physics \cite{AshcroftMermin}. Unavoidable disorder in periodic media will strongly alter the transport of electrons, spins, phonons or photons, ultimately resulting in the breakdown of transport, known as Anderson localization \cite{Anderson1958, Altshuler1991, Sheng1995, Akkermans2007, Lagendijk2009}. For ideal perfectly periodic systems the density of states (DOS) has a cusp at the band edge, known as the Van Hove singularity. In real systems, however, the Van Hove singularity is smeared out by disorder. For doped semiconductors and superconductors it is known that because of disorder the ensemble of localized states in the band gap forms a ``tail'' in the density of states that decays away from the band edge, known as the Lifshitz tail \cite{Lifshitz1964, Elliott1974, vanMieghem1992, Balatsky2006, Aizenman2011}.

Nanophotonic structures are excellent model systems to investigate the effects of disorder on band-edge phenomena in condensed matter physics in their purest form, since photons do not interact in linear media. Moreover, the availability of strong scatterers and nanofabrication methods allow for control over lengths much smaller than the wavelength. With techniques like near-field scanning optical microscopy (NSOM) it is possible to directly measure the wavefunction with sub-wavelength precision and high energy resolution \cite{Novotny2006}, a feat notoriously difficult in solid-state systems. Here we concentrate on photonic-crystal waveguides \cite{Krauss2007, Joannopoulos2008, Lund-Hansen2008, Falain2010}, which are formed by a two-dimensional (2D) photonic-crystal slab with a line defect that guides light with strong dispersion essential for slow light and enhanced light-matter interactions. Such a waveguide can be considered as a 1D system, in which the Van Hove singularity at the band edge of the guided mode is a divergence in the DOS \cite{John1994, Li2001}. It is expected that intrinsic disorder in photonic-crystal waveguides removes the Van Hove singularity, forming a mixed band filled with both extended and Anderson-localized modes \cite{John1987, Savona2011}. Indeed, localized states in photonic-crystal waveguides have been observed near the band edge \cite{Topolancik2007, Topolancik2009, Thomas2009, Sapienza2010, Garcia2010, Spasenovic2011}. However, no study of the density of states in the band gap has been made to elucidate the removal of the Van Hove singularity and formation of a Lifshitz tail.

\begin{figure}[t]
  \includegraphics[width=8.6 cm]{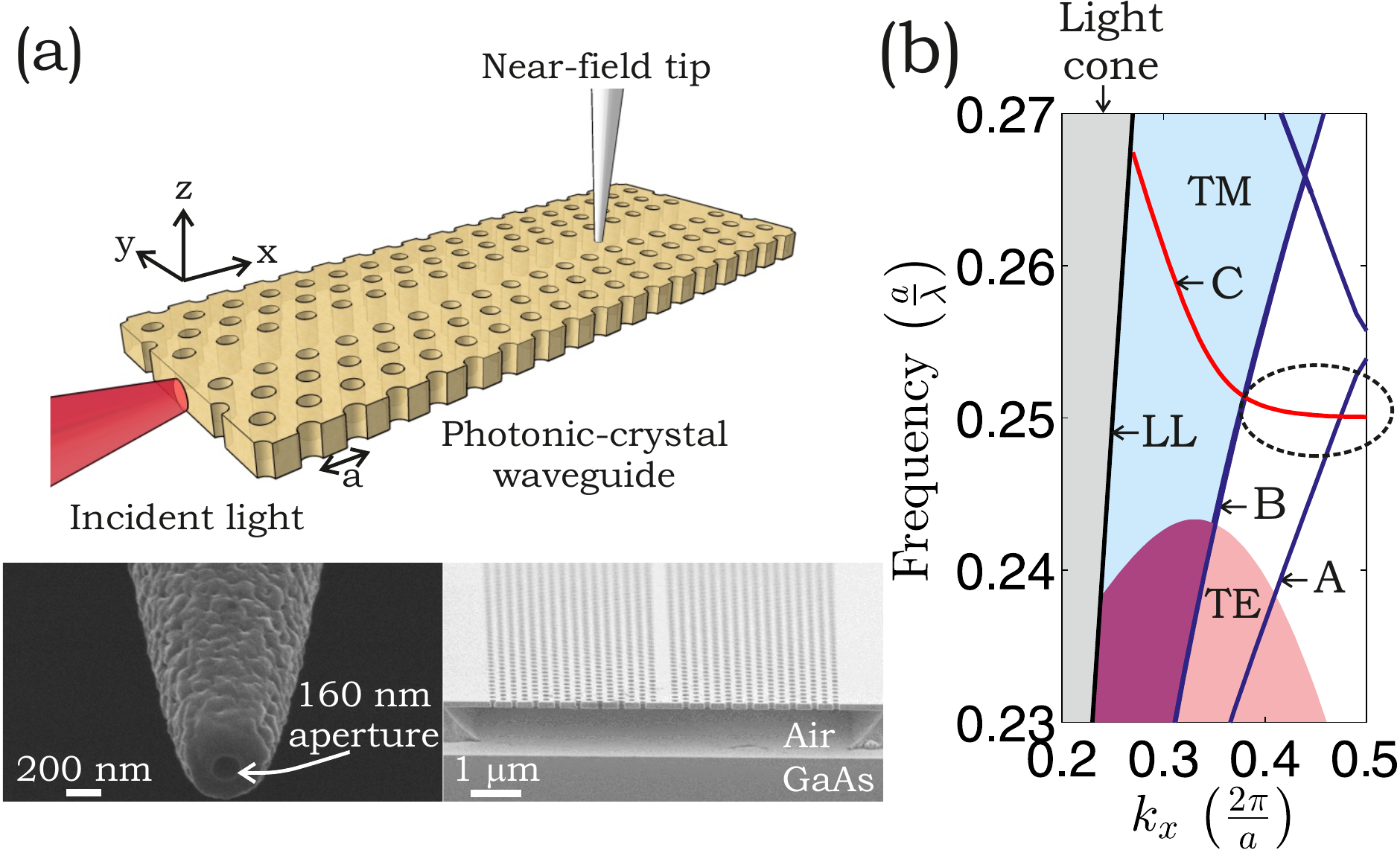}
\caption{\textit{(color online)} $(a)$ Laser light is side-coupled on a GaAs photonic-crystal waveguide. Light propagation is studied with a near-field tip. SEM images are shown of the coated near-field tip (left) and the waveguide (right). $(b)$ Calculated band structure showing both TE-like (red) and TM-like (blue) guided modes in the 2D bandgap for TE-like modes for ($\frac{r}{a}=0.303$, $\frac{h}{a}=0.67$). The encircled area is the main focus of this Letter. Here, the TE-like waveguide mode C becomes flat at the band edge, leading to a Van Hove divergence in the DOS. The black diagonal line represents the light line $(LL)$.}
\label{fig1}
\end{figure}

In this Letter we present near-field measurements of localized and extended modes in GaAs photonic-crystal waveguides. Our polarization-sensitive, frequency-resolved measurements demonstrate the intriguing coexistence of Anderson-localized TE-like modes with propagating TM-like modes which extend over the entire waveguide. These pumped extended modes excite the localized modes, revealing a unique combination of ballistic and localized light. We use phase-sensitive NSOM, because of its ability to probe evanescent fields and measure the dispersion relation  \cite{Spasenovic2011, Gersen2005,  Louvion2006, Mujumdar2007, Ha2011, Huisman2011, Riboli2011}. In the experimentally measured band structure the localized modes are seen to smear out the band edge. From the band structure the DOS is experimentally reconstructed, explicitly demonstrating the absence of the Van Hove singularity and a first direct observation of a Lifshitz tail.

Figure \ref{fig1}$(a)$ illustrates our experiment. A continuous-wave laser (Toptica DL pro 940) with a tunable wavelength $\lambda$ between $907$ and $990$ nm and a linewidth of $0.1$ MHz is side-coupled on a cleaved end facet of a GaAs photonic-crystal waveguide (right SEM image) with an objective (NA=0.55). The incident light is polarized with an angle of approximately $45^o$ with respect to the normal of the waveguide to excite both TE-like and TM-like modes. The field pattern is collected approximately $200$ \micro m away from the coupling facet using an aluminum coated near-field tip with an aperture of $160 \pm 10$ nm (left SEM image). We perform phase-sensitive NSOM using heterodyne detection \cite{Balistreri2000}.

The photonic-crystal waveguide consists of a $1$ mm long photonic-crystal slab with holes forming a triangular lattice with pitch $a=240$ nm, normalized hole radius of $\frac{r}{a}=0.309$, and slab thickness $h=160$ nm. A row of missing holes forms the W1 waveguide. Details on sample fabrication can be found in Refs. \onlinecite{Sapienza2010, Garcia2010}, where Anderson localization was demonstrated for waveguides fabricated under identical conditions. Figure \ref{fig1}$(b)$ shows the calculated band structure along the propagation axis for such a photonic-crystal waveguide \cite{mpb}. The blue and red bands describe modes that are guided by the line defect for TM- and TE-polarized light, respectively. The blue and pink areas mark continuums of modes propagating in the surrounding photonic crystal for TM- and TE-polarized light respectively, which overlap at the purple area. We concentrate on modes $(A)$ and $(B)$ that are TM-like, and mode $(C)$ that is TE-like. Intrinsic disorder causes Anderson localization in the slow-light regime of mode $(C)$ (near $k_x=0.5$, marked by the ellipse) \cite{Topolancik2007, Thomas2009, Garcia2010}, where the dispersion relation flattens and the optical density of states (DOS) ideally diverges.

\begin{figure}[t]
  \includegraphics[width=8.6 cm]{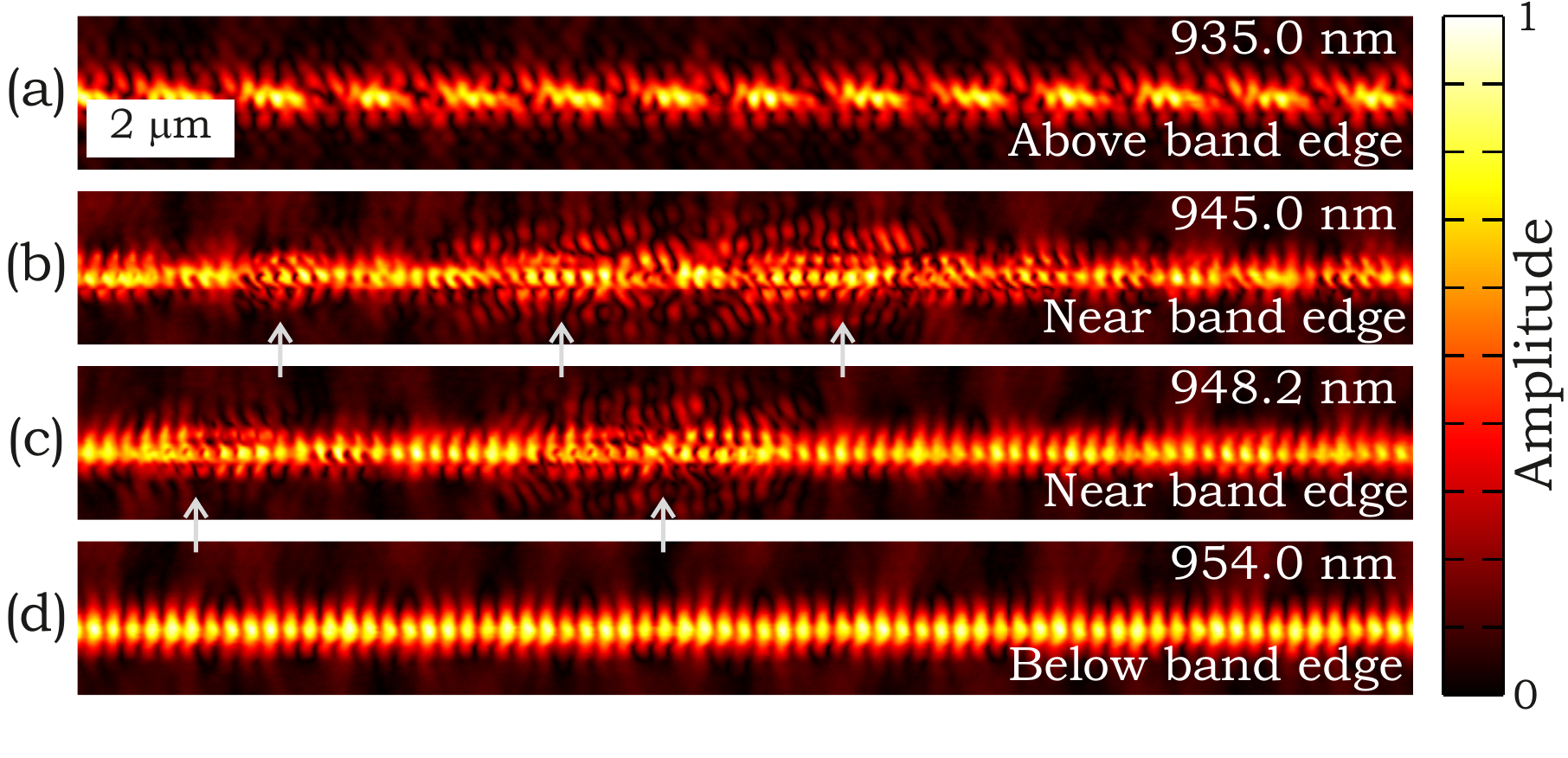}
\caption{\textit{(color online)} (a-d) Amplitude for a photonic-crystal waveguide measured at different frequencies above, near, and below the band edge. Propagating extended Bloch modes are observed in (a) and (d), Anderson-localized random modes are found in (b,c). The arrows mark localized states.}
\label{fig2}
\end{figure}

Figure \ref{fig2} presents measured near-field amplitudes at different excitation frequencies. Periodic beating patterns are observed above $(a)$ and below $(d)$ the band edge of TE-mode $(C)$. Spatial Fourier transforms \cite{Gersen2005} and Bloch-mode reconstruction \cite{Ha2009} confirm that these patterns are completely described by a superposition of Bloch modes \cite{Huisman2011}. For $907 \leqslant \lambda \leqslant 942.4$  and $954.9 \leqslant \lambda \leqslant 990\,$nm similar patterns of extended Bloch modes were observed to that of $(a)$ and $(d)$, respectively.

\begin{figure}[t]
  \includegraphics[width=6.6 cm]{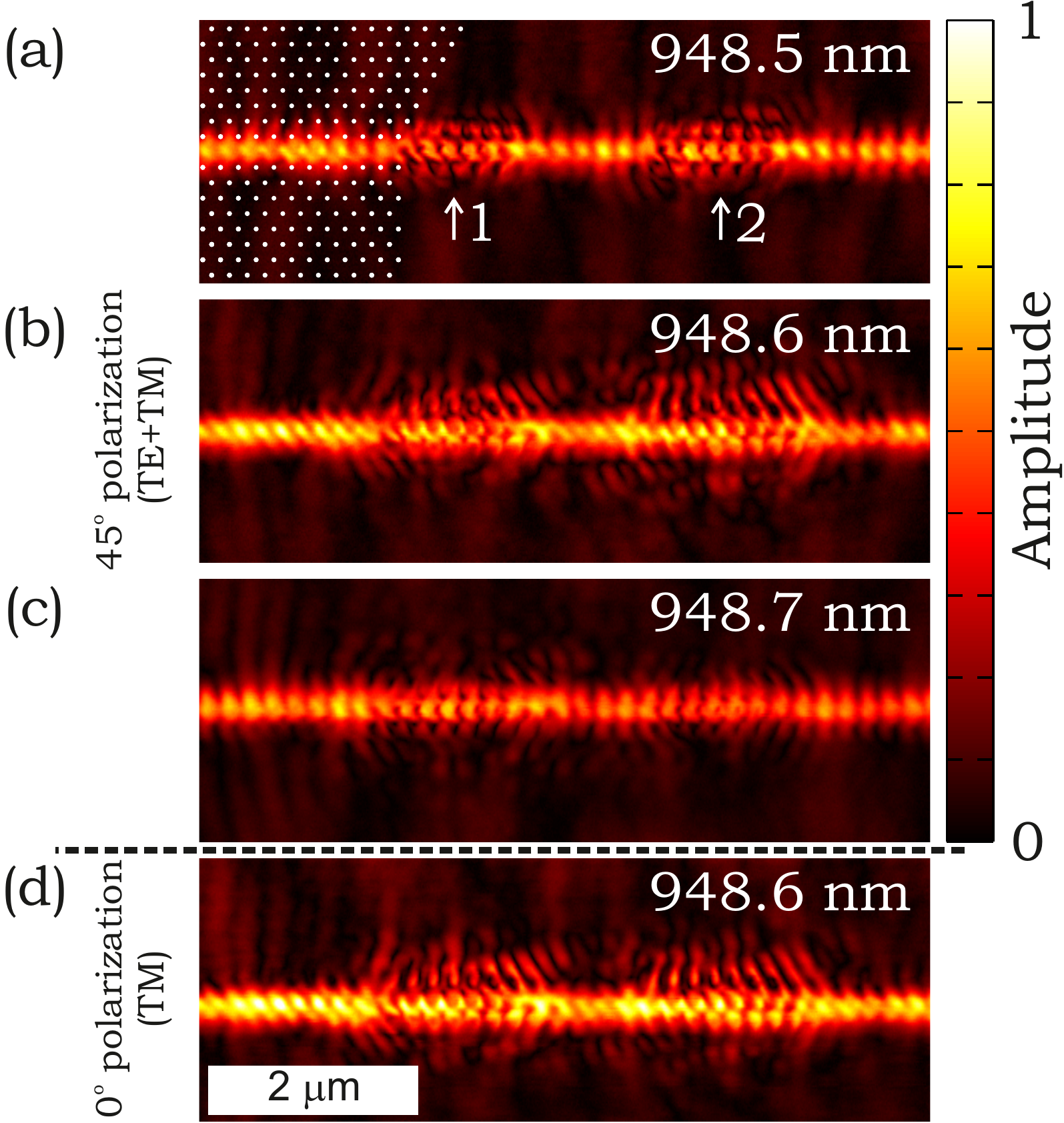}
\caption{\textit{(color online)} Amplitude for two coupled localized modes (arrow 1 and 2) for 3 wavelengths separated by $0.1$ nm for two incident polarizations with respect to the pores ($\approx 45^o$ $(a-c)$ and $0^o$ $(d)$). The amplitudes are normalized to the amplitude of light propagating along the surface of the structure. The white circles in $(a)$ indicate the location of the holes. }
\label{fig3}
\end{figure}

Figures \ref{fig2}$(b,c)$ are measured near the band edge, corresponding to frequencies in the range where the periodic patterns are perturbed by standing-wave field patterns (marked by arrows) that can extend up to approximately $3a$ into the surrounding crystal. These perturbations are the localized modes expected at the band edge. We have verified that these localized modes occur at random locations along the waveguide only within a wavelength range of $942.4-954.9\,$nm near the band edge of mode $(C)$. We have confirmed for three waveguides with different $\frac{r}{a}$ that the frequency range where such localized modes occur follow the shift of the predicted band edge, consistent with previous observations \cite{Topolancik2007}. The extended field patterns in the surrounding photonic crystal indicate high field amplitudes in the center of the waveguide. The observed maximum amplitude is likely quenched by the presence of the near-field tip. From spatial Fourier transforms we know that the periodic background is formed by TM-like modes $(A,B)$, which contain most of the field energy \cite{Huisman2011}. The observed localized modes agree well with calculated profiles of localized modes \cite{Savona2011, Smolka2011}. Moreover, from Bloch-mode reconstruction \cite{Ha2009} we know that they are not a superposition of Bloch modes. Clearly, from Fig. \ref{fig2} $(b-c)$ it can be deduced that the presence of the localized modes is strongly wavelength dependent. We have verified that the localized modes indeed have a narrow linewidth, see Fig. \ref{fig3}$(a-c)$ and \cite{LinewidthNote}.

We observe Anderson-localized modes far along the waveguide, where the intensity should naively be vanishingly small. Therefore we identify how the modes are excited. The near-field patterns of Fig. \ref{fig3}$(a-c)$ were obtained with an incident polarization angle of approximately $45^o$ with respect to the normal of the photonic-crystal waveguide to excite both TE- and TM-like modes. Fig. \ref{fig3}$(d)$ shows the near-field pattern at $\lambda=948.6$ nm when the incident polarization angle is $0^o$ to only excite TM-like modes. We observe an identical field pattern as in Fig. \ref{fig3}$(b)$. From spatial Fourier transforms only TM-like Bloch modes were identified, but the localized modes remained observed.
We conclude that we are detecting a subset of localized modes at the band edge for the TE-like mode $(C)$ that becomes excited by the TM-like propagating mode $(A)$, consistent with previous interpretations \cite{Topolancik2009}. We anticipate that the weak coupling between localized modes and ballistic light offers an opportunity to address, manipulate, and read-out light-matter interactions with localized modes deep inside a material.

\begin{figure}[t]
  \includegraphics[width=8.6 cm]{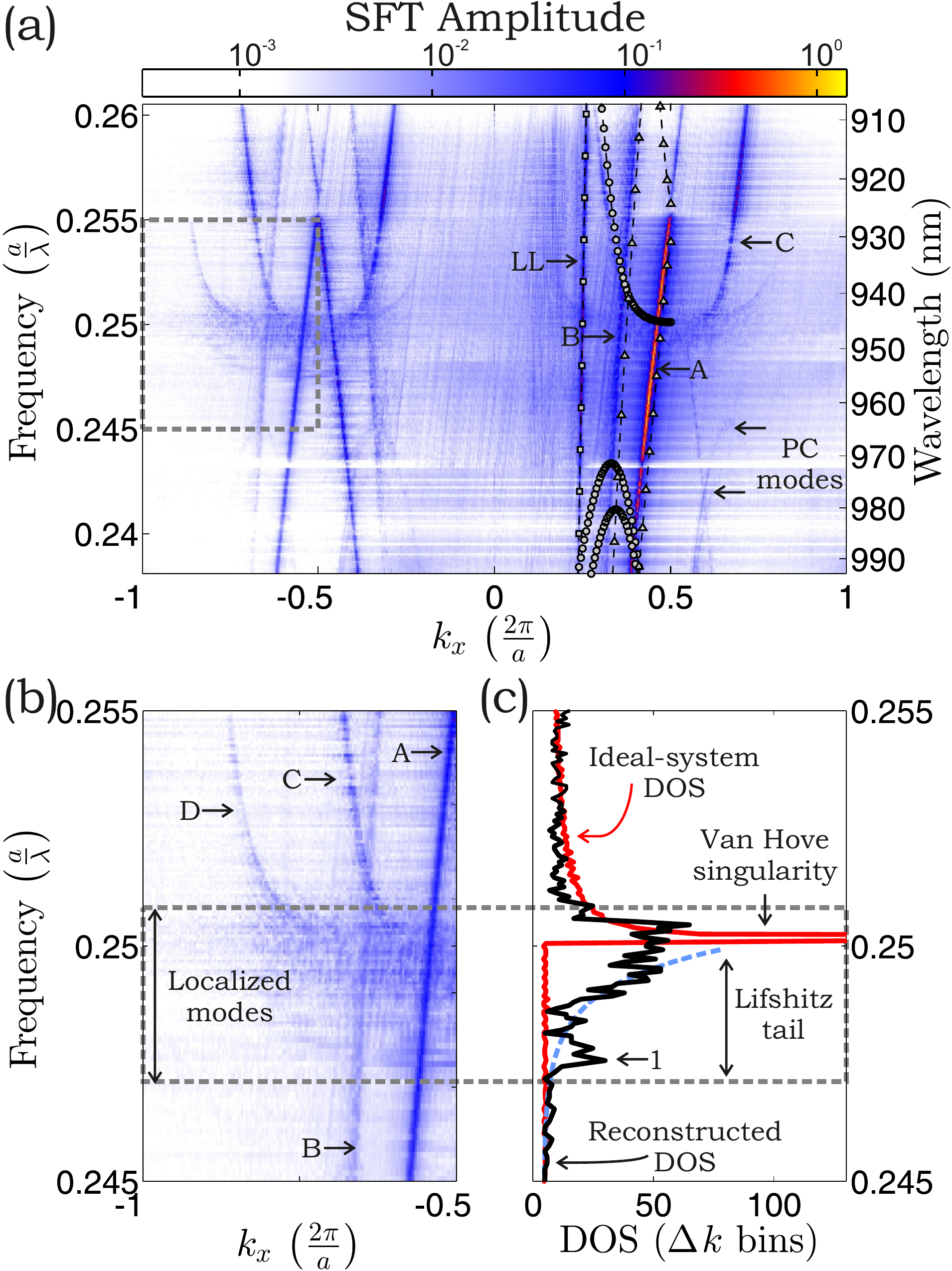}
\caption{\textit{(color)} $(a)$ Experimentally reconstructed band structure obtained from near-field patterns. The color represents the amplitude of the spatial Fourier (SFT) coefficients $\vert S_{\rm{int}}(k_x, \omega)\vert$. The black lines are fitted modes of Fig. \ref{fig1}$(b)$. The gray rectangle is highlighted in $(b)$. $(b)$ Localized modes are represented by the smeared out band edge. $(c)$ Experimentally reconstructed DOS (black) compared with a fitted DOS (red) for an ideal non-disordered waveguide. The single exponential fit (blue dashed) is a guide to the eye for the Lifshitz tail.}
\label{fig4}
\end{figure}

We have collected near-field patterns over a wide spatial range of $x=73$ \micro m and $y=2.5$  \micro m for a whole range of reduced laser frequencies $\omega=\frac{a}{\lambda}$ and obtained their spatial Fourier transform $S_{\rm{int}}(k_x,y, \omega)$. After normalization \cite{SFT} this resulted in the experimentally reconstructed band structure shown in Fig. \ref{fig4}$(a)$ \cite{Gersen2005}. Between $0 < k_x < 0.5$ the calculated folded band structure of Fig. \ref{fig1}$(b)$ is overlapped as symbols, showing good agreement with the maxima in $S_{\rm{int}}(k_x, \omega)$. One expects that Bloch harmonics repeat every Brillouin zone ($k_x=k_0+n \cdot a$, $n \in  \mathbb{Z}$) and are symmetric around the Bragg conditions $(k_x=n+0.5$, $n \in  \mathbb{Z}$). The identification of the modes as TE-like (circles) and TM-like modes (triangles) was confirmed by rotating the polarization of the incident light. We identify a narrow stopgap for mode $(A)$ at $\omega_s=0.2555$ ($k=-0.5,0.5)$, where the character of the modes changes. As a consequence, the overlap with the excitation beam changes and hence the spatial Fourier transform amplitudes of the modes change abruptly around $\omega\approx\omega_s$. The band edge for TE-polarized mode $(C)$ is located at $ \omega = 0.250$ $(k=-0.5,0.5)$.

In Fig. \ref{fig4}$(b)$ a zoom-in of the band structure is shown near the band edge of mode $(C)$ between $-1<k_x<-0.5$. A most intriguing feature is that our measurements do not show a sharp band edge, but a softened cutoff, in agreement with predictions of Ref. \cite{Savona2011}. The smearing out of the cutoff in the range between $0.247 < \omega < 0.251$ is caused by localized states, and starts to appear at a group index of $n_g(\omega = 0.251)=37$. No localized modes are observed outside this cutoff region. The width of the blurred cutoff $\Delta \omega = 0.004$ is a measure for the amount of disorder. If variations in the hole size and position are the dominant source of disorder \cite{Savona2011, Koenderink2005}, this would indicate a standard deviation of $\sigma=0.004 a$ in the hole positioning and hole radius, in reasonable agreement with sample characterization. The group index at which we first observe localized modes is consistent with our previous estimations \cite{Sapienza2010}. Mode $(D)$ is not a Bloch mode, but likely caused by an \textit{Umklapp} process between modes $(A)$ and $(C)$.

Figure \ref{fig4}$(c)$ shows the main result of this Letter: We have reconstructed the DOS from the bands in Fig. \ref{fig4}(b). For each $\omega$ we have calculated the number of $k_x$-bins that satisfy $\vert S_{\rm{int}}(k_x, \omega) \vert > q$, with $\Delta k_x = 0.0033$ and threshold $q=0.08$. This measure for the DOS is shown in Fig. \ref{fig4}$(c)$ (black). The shape of the reconstructed DOS is not very sensitive to the exact value of $q$.
We have applied a similar sampling method to the calculated band structure of Fig. \ref{fig1}$(b)$ to modes $(A, B, C)$, representing the calculated DOS of an ideal periodic waveguide (red). This DOS is scaled to have the same value as the experimental DOS between $0.255 > \omega > 0.254$. Both the experimental and calculated DOS are approximately constant between $0.255 > \omega > 0.251$ and for $0.247>\omega$. For these frequency ranges no localized modes are observed. Note the contribution of the TM-like modes which lead to a finite DOS for $0.247>\omega$. Between $0.251 > \omega > 0.250$ both the experimental and calculated DOS increase rapidly. The DOS of an ideal periodic system diverges to infinity as  $\rho(\omega) \propto (\omega-\omega_{\rm{gap}})^{-\frac{1}{2}}$, forming the Van Hove singularity~\cite{John1994, Li2001}. The experimental DOS follows this increase, until it saturates at $\omega=0.2505$. In this range the first localized modes are observed. Although the single peak at $\omega=0.2476$ (arrow 1) belongs to a single localized mode, it is unclear whether the sharp features in the reconstructed DOS are dominated by individual localized states, since localized states are present at every $\omega$ setting in this range. In future this could be answered by reconstructing band structures with smaller $\Delta \omega$. It is clear, though, that the Van Hove divergence is absent, in agreement with computations by Savona \cite{Savona2011}. In the band gap for $0.250 > \omega > 0.247$ the experimental and the calculated DOS significantly differ: The calculated DOS is constant, whereas the experimental DOS slowly decays away from the band edge, forming the Lifshitz tail known from solid-state systems \cite{Lifshitz1964, Elliott1974, vanMieghem1992, Balatsky2006, Aizenman2011}. Our present data do not allow to draw conclusions about the exact shape of the tail, which is debated in literature, but shows the possibility of addressing this issue in future experiments.

We have reported near-field measurements of localized modes using phase-sensitive NSOM; the experimentally obtained band structure reveals how the localized states perturb the band edge. Ensemble averaging by measuring band structures of different parts of the sample should smoothen the envelope of the reconstructed DOS and will give the possibility to quantitatively study the shape of the Lifshitz tail. We also predict that the Lifshitz tail should appear when the DOS is directly probed by studying emission of embedded quantum dots \cite{Lund-Hansen2008, WangLeistikow2011}.


We thank J. Bertolotti, D. Dikken, D. Garc\'ia, L. Kuipers, and H. Thyrrestrup for stimulating discussions, and C. Harteveld, J. Korterik and F. Segerink for technical support. This work was supported by FOM and NWO-Nano.


\begin{thebibliography}{8}
\bibitem{SFT}
We have calculated the SFT $S(k,y, \omega)$ for each line parallel with the line defect. We define $S_{\rm{int}}(k, \omega)=\int  \vert S(k,y, \omega) \vert dy$. All $S_{\rm{int}}(k, \omega)$ are normalized to have the same average value for $-3.5 < k_x < -1.8$, which is dominated by $\frac{1}{f}$ noise.
\end{thebibliography}
\end{document}